\documentclass{webofc}
\usepackage[varg]{txfonts}
\usepackage{array}
\usepackage{longtable}
\usepackage{appendix}

\usepackage{natbib}

\usepackage{array}
\usepackage[table]{xcolor}% http://ctan.org/pkg/xcolor

\newlength\mylength
\newcolumntype{C}[1]{>{\centering\arraybackslash}p{#1}}

\begin{document}

\title{CRIU - Checkpoint Restore in Userspace for computational simulations and scientific applications}

\author{\firstname{Fabio} \lastname{Andrijauskas}\inst{1}\fnsep\thanks{\email{fandrijauskas@ucsd.edu}} \and
\firstname{Igor} \lastname{Sfiligoi}\inst{1}\fnsep\thanks{\email{isfiligoi@sdsc.edu}} \and
\firstname{Diego} \lastname{Davila}\inst{1}\fnsep\thanks{\email{didavila@ucsd.edu}} \and
\firstname{Aashay} \lastname{Arora}\inst{1}\fnsep\thanks{\email{aaarora@ucsd.edu}} \and
\firstname{Jonathan} \lastname{Guiang}\inst{1}\fnsep\thanks{\email{jguiang@ucsd.edu}} \and
\firstname{Brian} \lastname{Bockelman}\inst{2}\fnsep\thanks{\email{bbockelman@morgridge.org}} \and
\firstname{Greg} \lastname{Thain}\inst{2}\fnsep\thanks{\email{gthain@cs.wisc.edu}} \and
\firstname{Frank} \lastname{Würthwein}\inst{1}\fnsep\thanks{\email{fkw@ucsd.edu}}
}

\institute{SDSC - UC San Diego, MC 0505 | 9500 Gilman Drive | La Jolla, CA 92093-0505 \and Morgridge Institute for Research | 330 N Orchard Street Madison WI}

\abstract{
Creating new materials, discovering new drugs, and simulating systems are essential processes for research and innovation and require substantial computational power. While many applications can be split into many smaller independent tasks, some cannot and may take hours or weeks to run to completion. To better manage those longer-running jobs, it would be desirable to stop them at any arbitrary point in time and later continue their computation on another compute resource; this is usually referred to as checkpointing. While some applications can manage checkpointing programmatically, it would be preferable if the batch scheduling system could do that independently. This paper evaluates the feasibility of using CRIU (Checkpoint Restore in Userspace), an open-source tool for the GNU/Linux environments, emphasizing the OSG’s OSPool HTCondor setup. CRIU allows checkpointing the process state into a disk image and can deal with both open files and established network connections seamlessly. Furthermore, it can checkpoint traditional Linux processes and containerized workloads. The functionality seems adequate for many scenarios supported in the OSPool. However, some limitations prevent it from being usable in all circumstances.
}

\maketitle

\section{Introduction} 

While executing scientific applications, it can be required to stop the process due to hardware problems or even end-of-life job problems. In that case, some applications can create files to save their current state from being loaded on a restoring process later. However, most of the software does not have this kind of feature. Some applications can create checkpoints; however, this process of dumping and loading software is a complex task due to the number of systems to control and the software's ability to control all the use case scenarios. This is a ''long dream`` of high performance computing and high throughput computing. Even more, saving and starting an application again can save money and time. These scientific applications have been used to discover new materials \cite{COLUCI2023171}, find black holes \cite{weitzel2017data}, and many others.  

OSG \cite{Pordes_2007} provides a distributed high throughput computing environment where campus research organizations can use their resources. This federation includes computing, data, and storage resources. The combination of HTCondor and Glidein Workflow Management System (GlideinWMS \cite{5171374}) provides access to computational resources. When more resources are required, HTCondor job execution daemons (aka glidein pilots) are submitted to the computer resources at the Glidein Factory (GF) sites; HTCondor and GlideinWMS are the base of OSPool. The OSPool is a computing capacity source accessible to any researcher affiliated with a US academic institution. Capacity is allocated following a Fair-Share policy.

CRIU (Checkpoint Restore in Userspace, pronounced kree-oo) is a tool for checkpointing and restoring applications in GNU/Linux environments \cite{widjajarto2021live,10.1145/3357526.3357542}. With CRIU, it is possible to stop an application, save the working memory on disk, and restore the state later. One project that can use the CRIU functionality is the OSPool. This work aims to create use cases to test CRIU on a high throughput and high performance computing environment and the OSPool to check how these use cases can take advantage of CRIU features. 

\section{Test setup}

To test CRIU's features, we created a list of use cases related to computational simulations and scientific applications and used two CRIU versions. One was the most updated version (3.17.1), and the second was a branch with non-root operations support \cite{WinNT}. This approach was necessary because scientific applications typically use an unprivileged account. The procedure was simple: executing a \textit{dumping} and \textit{restoring} CRIU command and checking the software output. After the restoring process, the system and kernel logs are inspected. We also used a well-known scientific application to emulate more complex scenarios and each case addresses a unique situation. In simple code scenarios, the objective was to create a simple and easy code to be debugged if necessary. 

The operation of dumping and loading software is very straightforward. To dump a software using a PID 9191: \texttt{criu dump --shell-job -t 9191}, to load an application is \texttt{criu --shell-job  load .}. Due to the necessity to load files, control network connection, and others, sometimes CRIU requires more parameters.

\section{Tests and results}

The goal with the use cases was to cover simple scenarios, such as in basic C code, until more complex scenarios related to networks and others. All the use cases and the results are in \url{https://path-cc.io/GIL/criu_checkpoint_restore_userspace/}. Table \ref{tbl1} shows an overview of all tests using CRIU. 

\begin{center}
\begin{table}

\begin{tabular}{ | m{13em} | m{4cm}| m{4cm} | } 
  \hline
  Tests & CRIU 3.17.1 & CRIU Branch non-root \\ 
  \hline
   Simple serial application & \color{teal}Working  & \color{teal}Working \\ 
  \hline
   Pthreading and forking & \color{teal}Working & \color{teal}Working \\ 
  \hline
   Applications with open files & \color{teal}Working & \color{teal}Working \\ 
  \hline
   Applications running in containers & \color{yellow} Partially working & \color{yellow} Partially working \\ 
  \hline
   Checkpointing while running inside a container runtime & \color{red} Not working & \color{red} Not working \\ 
    \hline
   CPU-specific optimizations & \color{teal}Working & \color{teal}Working \\ 
  \hline
   Applications using GPUs  & \color{red} Not working & \color{red} Not working \\ 
  \hline
   Network applications & \color{yellow} Partially working & \color{yellow} Partially working \\ 
  \hline
   Network file system & \color{teal}Working & \color{teal}Working \\ 
  \hline
   Parallel application & \color{red} Not working & \color{red} Not working\\ 
  \hline
\end{tabular}
  \caption{Each test group using CRIU and the overwall results.}
  \label{tbl1}
  \end{table}
\end{center}

\newpage
%root or non-root
\subsection*{Simple serial application}

Some scientific applications are purely compute-heavy, e.g., Monte Carlo simulations. After reading their inputs and getting additional input arguments, e.g., random seeds, they keep computing and do not interact with the environment until the end, when the outputs are created. We create this scenario with a simple C program that computes $\pi$ and writes the result to the terminal and other software for Molecular Dynamics. The first test, a ``Simple C'' test, shows a perfect CRIU execution using the standard version and the version with non-root capabilities. The other test used a serial LAMMPS opening a basic input of Lennard-Jones simulation \cite{Thompson2022}. CRIU shows that it is possible to run/dump/load using a LAMMPS serial version with both CRIU versions. These two results show that CRIU can be used in simple scenarios. However, there are more complex scenarios related to more compute-heavy processing.

% explain tests
\subsection*{Pthreading and forking}

One critical scenario is the utilization of multiple processes or threads; these types of programming techniques are used in molecular dynamics and other types of simulations. CRIU supports checkpointing threads or forks. To test this, two C codes were used. The first software was a PThread code showing a sequence of numbers and creating four threads; the other program created one fork to ``print'' a sequence of numbers.  Saving and loading applications with forks and threads was possible using both CRIU versions (root and non-root).

\subsection*{Applications with open files}

Finding software that loads the input and writes the output in a file is possible. An example of these features can be found on LAMMPS. LAMMPS loads files, and the result is written on the disk. CRIU can load and unload an application that uses files. However, keeping the same file structure is required, meaning the ``directory tree'' should be created on the computer to restore the process. This could be complicated if it is not known a priori where the application writes the file.

\subsection*{Applications running in containers}
 
Many applications rely on containerization these days for ease of portability and reproducibility. HTCondor can run user applications inside a container runtime, e.g., apptainer (was singularity), and it is indeed the most frequent use case in the HTCondor pilot setup. We thus tested the simple C program running inside an un-privileged apptainer, mimicking what an HTCondor pilot does. CRIU could not checkpoint such a job by invoking it outside apptainer nor inside apptainer, using both CRIU versions.

We then repeated the test by invoking CRIU using root privileges, mimicking the behavior of HTCondor as the host batch system manager. Even with added privileges, CRIU failed to checkpoint the user job. Next, running as root, we tested if replacing apptainer with podman and docker. The result did not change; CRIU could not checkpoint the user jobs running in the container. That said, docker does support checkpointing but has to be initiated directly through the docker toolset. 

It is possible to use Docker and podman with CRIU. However, it is necessary to use an interface on Docker or podman to stop or start containers, i.e., \texttt{docker checkpoint create looper checkpoint1} \cite{criudocker}. Singularity does not have this interface. Podman and Docker both have an interface to work with CRIU.  

\subsection*{Checkpointing while running inside a container runtime}

All the above tests were performed inside a virtual machine application, which closely mimics the behavior of a bare-metal setup. However, Some resource providers have started offering containerized resources, e.g., Kubernetes-based, for HTCondor pilots instead. Thus, we tested launching the simple C program inside the containerized environment and invoking CRIU in the same environment. Checkpointing failed in this setup.

\subsection*{CPU-specific optimizations}

Compiling the application's code using special flags specific to a given CPU is a common practice to speed up scientific applications. Using our simple C code, we proved that CRIU can work in this scenario if the CPU family/type is the same across the checkpointing and restoring processes. The test software can only be restored on a computer with the same family/type of processor used to compile the software.

\subsection*{Applications using GPUs}

To test the GPU with CRIU, a GPU matrix multiplication code was used.  CRIU does not support GPU checkpointing; all the attempts failed.  In fact, CRIU documentation explains this on the repository \cite{criugpu}.

\subsection*{Network applications}

Several scientific applications can connect to different hosts. This network connection could be related to a data set transfer, a user interface, or process communication. One example is the Matlab. Network applications were tested in different ways, starting with simple send-and-receive messages using TCP and UDP coded in C; cases related to starting and stopping using CRIU all in the same machine work very well. However, it is only possible to restore once on the same machine. If it is required to stop the software and change the machine, CRIU can not load the application again. The behavior of just being able to load the application is related to the firewall configuration to keep the connections alive during the unload/load process. Another case is the network file system.

\subsection*{Network file system}

On the OSPool, we have several network file systems to provide data to the users, container images, libraries, and others. One example of this type of network file system is CernVM-File System (CVFMS) \cite{Boyer_2022}. We use a docker and a simple bash script to test this scenario. The docker container image was mounting a CVMFS repository, and a script read files from the CVMFS mounting. With CRIU, stopping and restoring one software is possible without losing access to a remote file system.

\subsection*{Parallel application}

The message-passing interface (MPI) is a technology used to run software across different machines using the network. Using a LAMMPS application with MPI support, we could not perform checkpointing with CRIU. The load process got ''hung`` on all the attempts using the two CRIU versions (root and non-root).

\section{Conclusion}

Checkpointing could solve the problem of long simulations that are essential for advancing science, which, unfortunately, often fail to fit within the time constraints of batch computing providers.
From the point of view of batch system administration, stopping and restarting an application process on a different machine could open the door to support preemption without incurring low efficiency due to CPU waste time. Finally, it would help save the applications' progress affected by unplanned maintenance.  

CRIU can provide several options to stop and restore applications, it is possible to control applications with multiple threads and processes, and it is possible to maintain network connections. CRIU supports containers using docker and podman. It requires a ``form'' of root access: sudo, SUID Bit, or Kernel capabilities, and to use Kernel capabilities requires a specific version of CRIU and Linux. Restoring a previously checkpointed process requires the same directory paths used during restoration as during checkpointing. Restoring a previously checkpointed process requires the same directory paths used during restoration as during checkpointing. There is no support for GPUs, and this is an excellent feature to be tested in the future.

From the OSPool point of view, CRIU can not be used due to the limitation of the ''container`` interface. The container itself can not checkpoint itself. To do that is required to use the interface between docker and CRIU. That prevents CRIU from being used on the OSPool. This is another desired feature to be implemented on CRIU.

\section{Acknowledges}

This material is based upon work supported by the National Science Foundation under Grant No. 2030508. Any opinions, findings, conclusions, or recommendations expressed in this material are those of the author(s) and do not necessarily reflect the views of the National Science Foundation.

\bibliography{t}

\begin{thebibliography}{11}

\bibitem{COLUCI2023171}
V.R. Coluci, F.~Andrijauskas, S.O. Dantas, in \emph{Modeling, Characterization,
  and Production of Nanomaterials (Second Edition)}, edited by V.K. Tewary,
  Y.~Zhang (Woodhead Publishing, 2023), Woodhead Publishing Series in
  Electronic and Optical Materials, pp. 171--187, second edition~edn., ISBN
  978-0-12-819905-3,
  \urlstyle{tt}\url{https://www.sciencedirect.com/science/article/pii/B9780128199053000087}

\bibitem{weitzel2017data}
D.~Weitzel, B.~Bockelman, D.A. Brown, P.~Couvares, F.~W\"{u}rthwein, E.F.
  Hernandez, \emph{Data Access for LIGO on the OSG}, in \emph{Proceedings of
  the Practice and Experience in Advanced Research Computing 2017 on
  Sustainability, Success and Impact} (Association for Computing Machinery, New
  York, NY, USA, 2017), PEARC17, ISBN 9781450352727,
  \urlstyle{tt}\url{https://doi.org/10.1145/3093338.3093363}

\bibitem{Pordes_2007}
T.O.S.G.E.B. on~behalf of~the Osg Consortium:Ruth~Pordes, D.~Petravick,
  B.~Kramer, D.~Olson, M.~Livny, A.~Roy, P.~Avery, K.~Blackburn, T.~Wenaus,
  F.~Würthwein et~al., Journal of Physics: Conference Series \textbf{78},
  012057 (2007)

\bibitem{5171374}
I.~Sfiligoi, D.C. Bradley, B.~Holzman, P.~Mhashilkar, S.~Padhi, F.~Wurthwein,
  \emph{The Pilot Way to Grid Resources Using glideinWMS}, in \emph{2009 WRI
  World Congress on Computer Science and Information Engineering} (2009),
  Vol.~2, pp. 428--432

\bibitem{widjajarto2021live}
A.~Widjajarto, D.W. Jacob, M.~Lubis, Bulletin of Electrical Engineering and
  Informatics \textbf{10}, 837 (2021)

\bibitem{10.1145/3357526.3357542}
R.S. Venkatesh, T.~Smejkal, D.S. Milojicic, A.~Gavrilovska, \emph{Fast
  In-Memory CRIU for Docker Containers}, in \emph{Proceedings of the
  International Symposium on Memory Systems} (Association for Computing
  Machinery, New York, NY, USA, 2019), MEMSYS '19, p. 53–65, ISBN
  9781450372060, \urlstyle{tt}\url{https://doi.org/10.1145/3357526.3357542}

\bibitem{WinNT}
A.R. GitHub, \emph{Criu non-root} (1999),
  \urlstyle{tt}\url{https://github.com/adrianreber/criu/tree/non-root}

\bibitem{Thompson2022}
A.P. Thompson, H.M. Aktulga, R.~Berger, D.S. Bolintineanu, W.M. Brown, P.S.
  Crozier, P.J. in~'t Veld, A.~Kohlmeyer, S.G. Moore, T.D. Nguyen et~al.,
  Computer Physics Communications \textbf{271}, 108171 (2022)

\bibitem{criudocker}
CRIU, \emph{Criu} (1999), \urlstyle{tt}\url{https://criu.org/Docker}

\bibitem{criugpu}
CRIU, \emph{Criu} (2018),
  \urlstyle{tt}\url{https://github.com/checkpoint-restore/criu/issues/534}

\bibitem{Boyer_2022}
A.F. Boyer, C.~Haen, F.~Stagni, D.R.C. Hill, in \emph{Lecture Notes in Computer
  Science} (Springer International Publishing, 2022), pp. 354--371,
  \urlstyle{tt}\url{https://doi.org/10.1007%2F978-3-031-07312-0_18}

\end{thebibliography}

\end{document}